\documentclass[aps,prl,twocolumn,amsmath,amssymb,floatfix, superscriptaddress, 10pt]{revtex4-2}
\usepackage{tabularx}
\usepackage{bm}
\usepackage{euscript}
\usepackage{epsfig,psfrag,subfigure}
\usepackage{graphicx}
\usepackage{color}
\usepackage{amsfonts}
\usepackage{exscale}
\usepackage{amsbsy}
\usepackage{multirow}
\usepackage[normalem]{ulem}
\usepackage[colorlinks=true,linkcolor=blue,citecolor=blue, urlcolor=blue,anchorcolor=blue]{hyperref}
\usepackage{tikz}
\usepackage{xcolor}
\usepackage{tikz}
\usepackage{physics}
\usepackage{amssymb}
\usepackage{mathtools}
\usepackage{amsmath}
\usepackage{dsfont}

\definecolor{xuzhi}{RGB}{210, 160, 20}

\begin{document}
\title{Superconductivity and competing orders in honeycomb $t$-$J$ model: interplay of lattice geometry and next-nearest-neighbor hopping}
\author{Zhi Xu}
\affiliation{School of Physical Science and Technology, ShanghaiTech University, Shanghai 201210, China}

\author{Hong-Chen Jiang}
%\email{hcjiang@stanford.edu}
\affiliation{Stanford Institute for Materials and Energy Sciences, SLAC National Accelerator Laboratory and Stanford University, Menlo Park, CA 94025, USA}

\author{Yi-Fan Jiang}
\email{jiangyf2@shanghaitech.edu.cn}
\affiliation{School of Physical Science and Technology, ShanghaiTech University, Shanghai 201210, China}

\begin{abstract}
We investigate the extended $t$-$J$ model on honeycomb lattices with next-nearest-neighbor (NNN) electron hopping $t'$ and superexchange coupling $J'=(t'/t)^2 J$ using large-scale density-matrix renormalization group (DMRG) simulations and slave-boson mean-field theory (SBMFT). By systematically varying $t'$ and cylinder geometries, our DMRG results reveal several competing phases with distinct charge and superconducting (SC) properties. On YC4-0 cylinders possessing bonds lying along $\vec{e}_y$ direction, the ground state of doped models exhibits pronounced quasi-long-range $d$-wave SC with coexisting armchair-oriented stripes (a-stripe) across a broad range of $t'$. Notably, the SC Luttinger exponent has a non-monotonic dependence on $t'$, showing an optimal  $t'_{op}\sim0.4$ for dominant SC. Conversely, XC cylinders host a competing long-range zigzag stripes phase without SC for $t'>0.5$, highlighting the role of boundary geometry in stabilizing distinct competing phases in DMRG. To elucidate the stability of all these competing phases in 2D limit, we employ SBMFT and identify the a-stripe as the stable configuration across most of phase diagram, with a transition to uniform nematic $d$-wave SC at large $t'$ for $\delta=1/8$. The combined results from two complementary approaches suggest a robust $t'$-induced SC phase that might remain stable in doped extended %2D honeycomb-lattice 
$t$-$J$ model on the honeycomb lattice.
\end{abstract}

\maketitle

%Introduction

Understanding physics emerging from doped Mott insulators is one of the central challenge in strongly correlated electron systems. As paradigmatic models incorporating electron correlation, the Hubbard model and its descendant, the $t$-$J$ model, have been extensively investigated since the discovery of cuprate superconductors. It is widely believed that these seemingly simple models on square lattice capture the complex interplay between superconductivity (SC) and competing charge/spin order in high-$T_c$ cuprates \cite{arrigoni2004mechanism, white2003stripes, berg2009striped, zheng2017, Ido2018, Boris2019, qin2022review, jiang2023density, yang2023recovery, jiang2023pair, Jiang2019,Xu2024,Gu2025,Christopher2025Superconductivity,Jiang2024hub,Xu2024Pair,Qu2024,Jiang2025Hub}.
As a counterpart to the square lattice, the honeycomb lattice Hubbard model has also garnered considerable interest.  This interest has recently been further amplified by discoveries of correlated states in moiré systems such as twisted bilayer graphene \cite{Li2009,Andrei2021,Cao2018insulator,Cao2018superconductivity} and tunable honeycomb-lattice Hubbard model realized in twisted MS$_2$ or MSe$_2$ (M=Mo, W) \cite{Yuan2018, Pan2020, Angeli2021, Pan2023, Ma2025, Wei2025}.
The non-interacting honeycomb model with nearest neighbor (NN) hopping hosts a Dirac semi-metal phase at half-filling. 
When the onsite Hubbard interaction $U$ exceeds a critical $U_c\sim 3.8$, the systems undergoes a metal-to-insulator transition into an anti-ferromagnetic (AFM) Mott phase \cite{Assaad2013, Otsuka2016, Yadav2023}.

Doping the system away from half-filling triggers a competition between hole kinetic energy and electron interactions, leading to unconventional phenomena that have drawn extensive interest. Several distinct candidates 
have been proposed as the ground states for doped honeycomb systems. 
In weak coupling, random phase approximation \cite{Xiao2016}, mean feild (MF) thoery \cite{Qi2020}, Quantum Monte Carlo (QMC) \cite{Pathak2010,Ma2011,Guo2024}, variational Monte Carlo (VMC) \cite{Li2022} and renormalization group approaches \cite{Honerkamp2008,Raghu2010,Nandkishore2012,Kiesel2012,Wang2012} consistently suggest the emergence of a $d+id$ SC phase. This phase is also observed in strong coupling regime via MF \cite{Ho2023,Jia2017,Zhong2015} and tensor product states (TPS) \cite{Miao2025,Gu2013,Xu2023}. In many works, $d+id$ SC phase competing or co-existing with other candidates include $s$-wave, $p+ip$ and $f$-wave pairing SC \cite{Cui2024,Uchoa2007,Faye2015,Gu2020,Kiesel2012,Xu2016}. Additionally, stripe-ordered state without SC is observed using DMRG, QMC and TPS \cite{Yang2021,Qin2022}, while recent DMRG study reports SC intertwined with stripe orders in the Hubbard model \cite{Peng2025}.
Exotic phases like pair-density-wave SC also emerges in spin-polarized honeycomb models \cite{Jiang2024}.
Experimentally, evidence of SC is reported in doped graphene \cite{Chapman2016,yang2024} and block material with honeycomb layers such as MgB$_2$, SrPtAs, FePSe$_3$ \cite{Nishikubo2011,Wang2018,Baskaran2002}.

The introduction of NNN hopping $t'$ and corresponding Heisenberg exchange $J'$ dramatically expands the phase diagram. At half filling, the $J$-$J'$ Heisenberg model hosts a nonmagnetic valence bond solid (VBS) phase for $J'/J\sim0.2-0.4$ \cite{Zhu2013,Albuquerque2011,Mezzacapo2012,Reuther2011,Mosadeq2011,Ganesh2013,Ferrari2017,Gong2013}, with a putative quantum spin liquid (QSL) phase sandwiched between AFM and VBS phases \cite{Cabra2011, Clark2011, Mukherjee2023, Gong2013}. 
The role of $t'$ on square lattice Hubbard models has been intensively investigated and appears crucial for stabilizing unconventional SC. While electron-doped ($t'>0$) systems show promising SC tendencies, the ground state of hole-doped ($t'<0$) cases remain controversial due to strong competition between CDW and SC orders \cite{Jiang2019,Xu2024,Xu2025asymmetry,Gu2025,Christopher2025Superconductivity,Jiang2024hub,Xu2024Pair,Qu2024,Jiang2025Hub}. This fundamental issue regarding the emergence of SC upon hole doping extends to honeycomb lattice models, where the intricate undoped phase diagram featuring AFM, QSL and VBS phases may further diversify emergent phases induced by doping.  However, the specific role of $t'$ in governing SC and its competition with CDW in doped honeycomb lattice $t$-$J$ models remains largely unexplored.

\begin{figure}[t]
    \centering
    \includegraphics[width=1\linewidth]{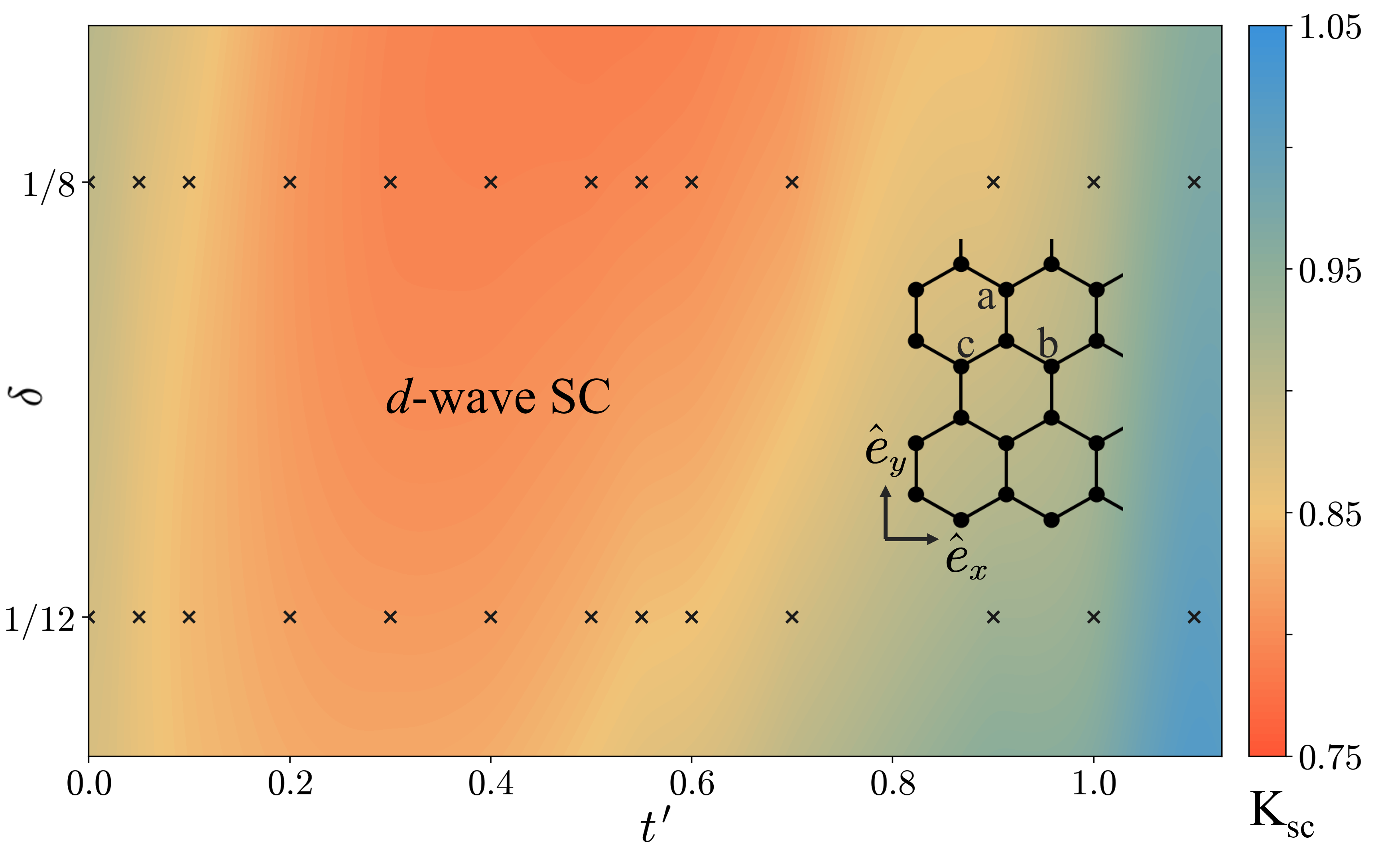}
    \caption{Quantum phase diagram of the $t$-$t'$-$J$-$J'$ model on YC4-0 cylinders with $t=2$ and $J=1$, featuring a SC phase characterized by dominant quasi-long-range SC and subleading CDW orders at $\delta=1/12$ and $1/8$. Color scale denotes the Luttinger exponent $K_{sc}$ of SC correlation, $\Phi\sim r^{-K_{sc}}$. Inset: Lattice geometry and bond labels of YC4-0 cylinders. }
    \label{fig1:phasediagram}
\end{figure}

{\it Principal results:}
In this work, we address this open question by systematically investigating the $t$-$t'$-$J$-$J'$ model on the honeycomb lattice using large-scale density-matrix renormalization group (DMRG) simulations and slave-boson mean-field theory (SBMFT). Our DMRG simulations on YC4-0 cylinders reveal a robust SC phase characterized by dominant quasi-long-range nematic $d$-wave SC correlation and sub-leading armchair-oriented stripe charge order (a-stripe) across a wide range of $t'$ and doping $\delta$, as summarized in Fig.~\ref{fig1:phasediagram}. Intriguingly, the SC exponent $K_{sc}$ of this phase exhibits a non-monotonically valley-shaped $t'$-dependence: it is enhanced in doped AFM regime at small $t'$, reaches a minimum $K_{sc}\sim 0.80$ (optimal SC) at $t'_{op} \sim 0.4$ $(t=2)$, and weakens yet maintaining dominant ($K_{sc}<1$) in the doped VBS regime for $t'$ up to $1.0$. 

The competition between SC and CDW orders shows striking geometry dependence on finite cylinders. The other candidate states, e.g. long-range zigzag-oriented stripe (z-stripe) and bidirectional CDW order, emerged on varied types of cylinders highlight the critical role of boundary geometry in stabilizing distinct phases. To resolve this finite-size effect, subsequent SBMFT calculations are employed to evaluate the fate of these candidate states in the thermodynamic limit. The resulting mean-field (MF) phase diagram (Fig.~\ref{fig5:MF-phase}) suggests that the a-stripe pattern observed on YC4-0 cylinder persists in the ground state over other CDW orders in 2D limit. For $\delta=1/8$, the MF phase diagram further reveals a transition from a dominant a-stripe phase at small $t'$ to a nematic $d$-wave SC phase without charge order when $t'>0.5$, consistent with the significant suppression of CDW correlation observed in DMRG. By combining two complementary methods, our results provide the first comprehensive evidence for NNN-hopping-enhanced SC in the extended honeycomb $t$-$J$ model.

% model and method
{\it Models and method:}
Using large-scale DMRG simulations \cite{White1992, Schollwöck2005}, we investigate the ground-state properties of the $t$-$t'$-$J$-$J'$ model on honeycomb lattices under various boundary conditions. The Hamiltonian is defined as: 
\begin{eqnarray}\label{Eq:hami}
    H&=& \mathcal{P} H_t \mathcal{P} + H_s ~ ,\nonumber \\
    H_t&=&\sum_{ij\sigma}( - t_{ij} \hat{c}_{i\sigma}^{\dagger}\hat{c}_{j\sigma}+h.c.) ~ ,\nonumber \\
    H_s&=&\sum_{ij}J_{ij}\left( \mathbf{S}_i\cdot \mathbf{S}_j-\frac{1}{4}\hat{n}_i\hat{n}_j \right) ~ ,
\end{eqnarray}
where $\hat{c}_{i\sigma}^{\dagger}$ ($\hat{c}_{i\sigma}$) creates (annihilates) a spin $\sigma$ electron at site $i$, $\mathbf{S}_i$ denotes the spin operator and $\hat{n}_i=\sum_{\sigma} \hat{c}_{i\sigma}^\dagger \hat{c}_{i\sigma}$ is the electron density operator. The Gutzwiller projection operator $\mathcal{P}$ enforce the no-double-occupancy constraint. The hopping integral $t_{ij}=t$ for NN bond and $t_{ij}=t'$ for NNN bonds. We adopt NN AFM coupling $J=1$ as the energy unit and fix $t=2$, corresponding to an effective Hubbard interaction of $U=4t^2/J=8t$. In present study, we mainly consider the $t'>0$ case and set the NNN AFM coupling $J'=(t'/t)^2 J$ following the super-exchange relation. 

% Figure 1: phase diagram, main results

Since the CDW and SC orders on finite cylinders are sensitive to the lattice geometry \cite{Talkachov2023,Xu2024Pair}, we perform DMRG simulations on three types of cylinders, i.e., YC4-0, XC8-0 and YC4-4 cylinders, as demonstrated in the Supplementary Materials (SM). These cylinders have periodic (open) boundary condition along the $\vec{e}_y$ ($\vec{e}_x$) direction, and X(Y) indicates that one of the three NN bonds lies along the $\vec{e}_x(\vec{e}_y)$ direction. YC4-$n$ cylinders consist of 4 zigzag chains along $\vec{e}_x$, with periodic connections shifted by $n$ sites along the zigzag chain. XC8-0 cylinders contains 4 armchair chains along $\vec{e}_x$. 
The system size is counted by the number of unit cells $L_x$ and $L_y$ along $\vec{e}_x$ and $\vec{e}_y$ directions, respectively. 
The doping concentration is $\delta=1-N_e/N$, where $N_e$ is number of electrons and $N=2\times L_x \times L_y$ is the total number of sites. In this work, we focus on $\delta=1/8$ and $1/12$ doped models with $N=384$. In DMRG simulation, we keep up to $m=24000$ block states and perform at least 80 sweeps to ensure the convergence. Additional details on numerical methods are provided in SM.

% Figure 2: SC correlation and charge density profiles
\begin{figure}[tb]
    \centering
    \includegraphics[width=1\linewidth]{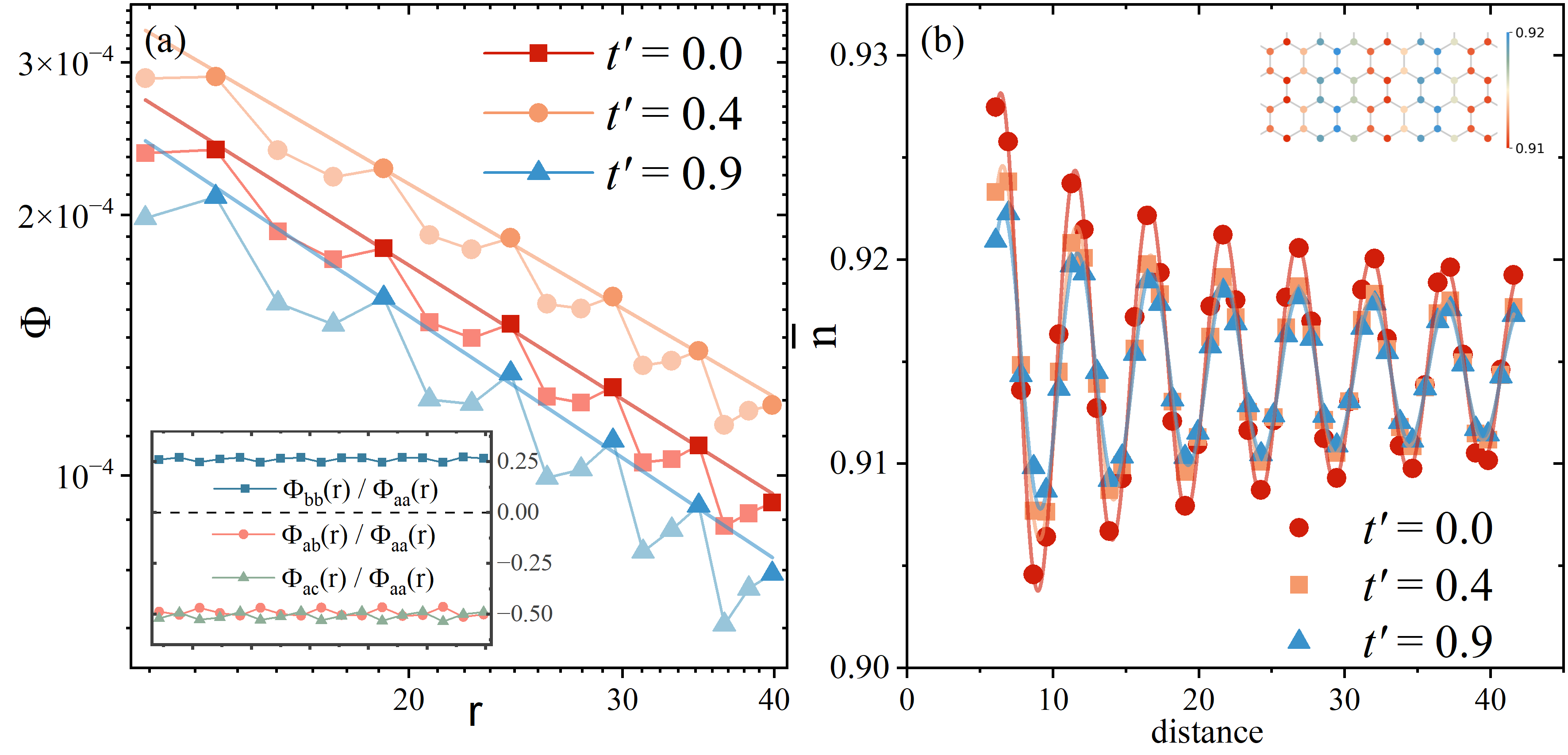}
    \caption{The SC and charge properties of $\delta=1/12$ doped model on YC4-0 cylinders. (a) The SC correlation $\Phi_{aa}$ measured at $t'=0.0$, $0.4$ and $0.9$, fitted to power-law functions $f(r)\sim r^{-K_{sc}}$ (solid lines). The transparent points far from the envelope are discarded in the fitting process. Inset: the ratios between different types of SC correlations. (b) The rung-averaged charge density $n(x)$ of $t'=0.0$, $0.4$ and $0.9$ models, fitted to the Friedel oscillation (solid lines). Inset: Bulk charge-density profile of $t'=0.0$ model.}
    \label{fig2:SCcorandcharge}
\end{figure}

{\it Robust SC phase on YC4-0 cylinder:}
We first investigate the ground-state properties of the model on YC4-0 cylinders. The SC properties are diagnosed through equal-time pair-pair correlation functions: 
\begin{equation}
    \Phi _{\alpha \beta}\left( r \right) =\left< \Delta _{\alpha,( x_0,y_0 )}^{\dagger} \Delta _{\beta,(x_0+r,y_0)} \right> ,
\end{equation}
where $\hat{\Delta} _{\alpha,(x,y)}^{\dagger} =\frac{1}{\sqrt{2}}\left( c_{\uparrow ,\left( x,y \right)}^{\dagger}c_{\downarrow ,\left( x,y \right) +\alpha}^{\dagger}-c_{\downarrow ,\left( x,y \right)}^{\dagger}c_{\uparrow ,\left( x,y \right) +\alpha}^{\dagger} \right)$ creates a spin-singlet pair on bond $\alpha$=a, b, c originating from site $(x,y)$. The reference site $(x_{0},y_{0})$ is positioned at $x_0\sim{\sqrt{3} L_{x}/{4}}$ and $r$ denotes the spatial separation along the $\hat{e}_x$ direction. Fig.~\ref{fig2:SCcorandcharge}(a) summarizes SC correlations for $\delta=1/12$ doped models with $t'=0.0 - 0.9$, which clearly show quasi-long-range decay behaviors characterized by
\begin{equation}
    \Phi_{\alpha \beta}(r)\propto r^{-K_{sc}}. 
\end{equation}
The extracted Luttinger exponents are $K_{sc}=0.88(2)$, $0.82(2)$ and $0.93(4)$ for $t'=0.0$, $0.4$ and $0.9$ models, indicating a dominant SC correlation with non-monotonic dependence of $K_{sc}$ on $t'$. 
While the $K_{sc}$ is nearly same for all types of $\Phi_{\alpha \beta}$ ($\alpha,\beta=a,b,c$) for a given $t'$, the relative magnitudes of SC correlations are dominant on $a$ bonds, $\Phi_{aa}(r) \sim 4 \Phi_{bb}(r) \sim 4 \Phi_{cc}(r)$, partially due to the broken $C_3$ symmetry on the cylinder. As illustrated in the inset of Fig.~\ref{fig2:SCcorandcharge}(a), SC correlations between different bond types present clear sign change, $\Phi_{aa}(r)\sim - 2\Phi_{ab}(r)\sim - 2\Phi_{ac}(r)$, suggesting that the pair symmetry is consistent with nematic $d$-wave.
Similar results are also observed in the $\delta=1/8$ cases (See SM for details). 

The charge density properties are determined by measuring the local density operator $\hat{n}(x, y)$. The density profile $\left< \hat{n}(x, y) \right>$ of $t'=0$ model in the inset of Fig.~\ref{fig2:SCcorandcharge}(b) reveals a characteristic stripe pattern breaking translation symmetry along $\hat{e}_x$ direction. Since the charge density is invariant along armchair direction, we denote it as ``a-stripe'' and calculate its rung average $n(x)=\sum_y{\left<\hat{n}(x,y)\right>}/L_{y}$ to analyze its spatial distribution. In all $t'=0.0 - 0.9$ cases, the ground states form half-filled a-stripe with two doped holes per CDW supercell on 4-leg cylinders. The wavelength of stripe is $\lambda=\sqrt{3}/4\delta$ along $\hat{e}_x$. Similar stripes with $\lambda=\sqrt{3}/L_y\delta$ ($1/L_y\delta$ unit cell) are also observed on $L_y=2$ and $3$ cylinders. 
At long distances, quasi-long-range CDW correlations are characterized by a power law functions with Luttinger exponent $K_c$. This exponent can be extracted from the charge density (Friedel) oscillations induced by cylinder boundaries \cite{White2002Friedel},
\begin{eqnarray}\label{Eq:CDW}
    n(x)=A \cos{(Qx+\phi)} x^{-K_{c}/2}+n_0,
\end{eqnarray}
where $Q=2\pi/\lambda$ denotes the ordering vector of CDW. $A$ and $\phi$ are nonuniversal amplitude and phase, respectively. From the fitting line illustrated in Fig.~\ref{fig2:SCcorandcharge}(b), we extract the exponents $K_c$ is $1.15(2)$ for the $\delta=1/12$ and $t'=0$ model.
Similar to pair correlations, density correlations maintain quasi-long-range behavior as $t'$ gradually increases. As shown in Fig.~\ref{fig2:SCcorandcharge}(b), while the wavelength of stripe remains nearly $t'$-independent, the decay exponents $K_c$ and amplitude of charge oscillation evolve with $t'$. 

% Figure 3: Luttinger exponents
\begin{figure}[tb]
    \centering
    \includegraphics[width=1\linewidth]{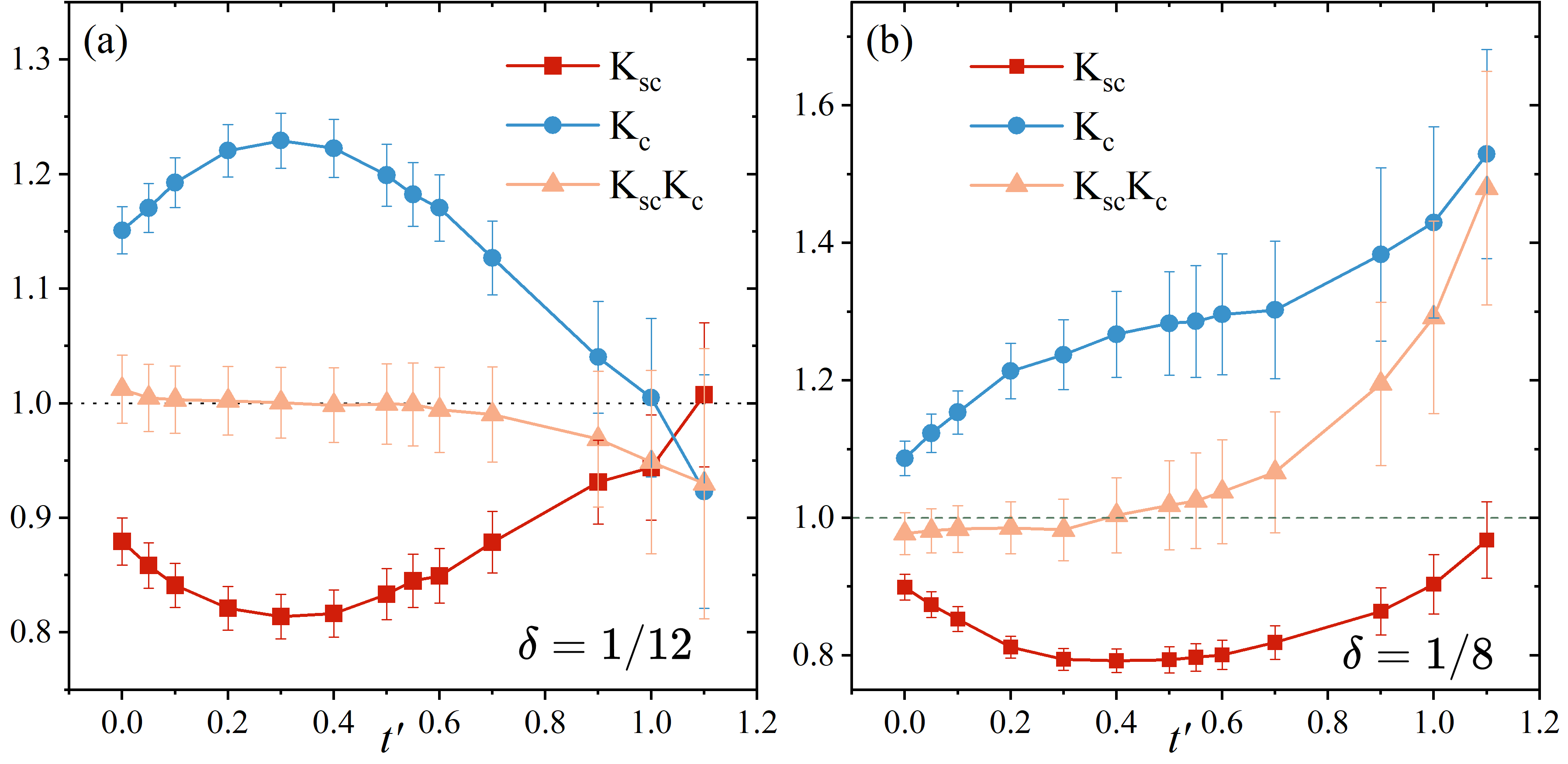}
    \caption{(a) The extracted Luttinger exponents $K_{sc}$ and $K_c$ of $\delta=1/12$ doped models with a series of $t'=0.0\sim 1.1$. Dominant SC correlations with $K_{sc}<1<K_{c}$ are observed for most of $t'$. (b) The exponents for the $\delta=1/8$ model. While SC correlations remain dominant across all $t'$, a significant suppression of CDW is observed at $t'>0.5$.}
    \label{fig3:exponent}
\end{figure}

The evolution of $K_{sc}$ and $K_{c}$ reveals a profound competition between SC and CDW orders mediated by $t'$. For $\delta=1/12$, the evolution of $K_{sc}$ in Fig.~\ref{fig3:exponent}(a) show enhanced SC correlation that being optimal ($K_{sc}\sim 0.82$) at $t'_{op}\sim 0.4$ in small $t'$ region, corresponding to a $J'\sim 0.02J$ that deep inside the AFM phase of the undoped phase diagram. 
Beyond this optimal $t'$, SC correlation weakens but remains dominant as indicated by $K_{sc} < 1 < K_{c}$ holding in most of $t'$ range. This valley-shape $t'$-dependence suggests that the spin fluctuations from AFM phase could play a paramount role in enhancing SC upon doping, which might be equally important as those arising from quantum critical point or QSL phase around $J'\sim 0.2J$. 
The exponent $K_{c}$ exhibits inverse trend to that of pair correlations as shown in Fig.~\ref{fig3:exponent}(a). The product of two exponents satisfy the relation $K_{c} K_{sc}=1$, consistent with a Luther-Emery liquid featuring a spin gap and single gapless bosonic mode.

For higher doping $\delta=1/8$, while the leading $K_{sc}$ still exhibits a similar valley-shaped dependence on $t'$ in Fig.~\ref{fig3:exponent}(b), the CDW order becomes significantly suppressed for $t'>0.5$ as evidenced by a rapid increase of $K_{c}$. The persistence of dominant SC correlations with $K_{sc}<1$ despite weakened stripe orders implies a tendency toward a nematic $d$-wave SC phase with uniform charge density driven by large $t'$.

% Figure 4: CDW in XC cylinder
\begin{figure}[tb]
    \centering
    \includegraphics[width=1\linewidth]{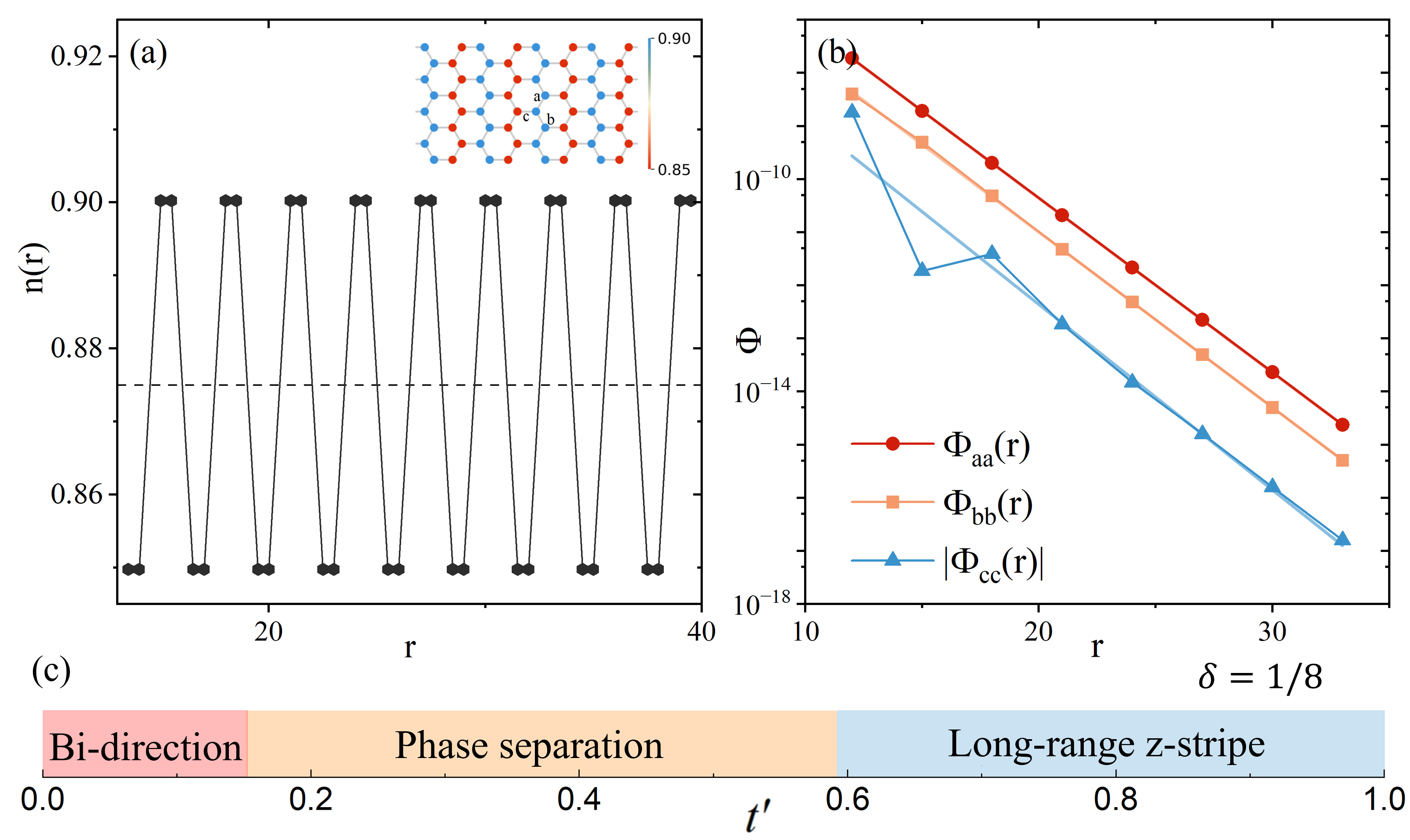}
    \caption{ The properties of the z-stripe phase on the XC8-0 cylinders. (a) The rung-averaged charge density $n(x)$ of the $t'=0.6$ models with $\delta=1/8$. (b) Three types of SC pair correlations measured in $t'=0.6$ model, fitted to exponential function $f(r)\sim e^{-r/\xi}$ with correlation length $\xi\sim 0.4$. (c) The $t'$ phase diagram of $\delta=1/8$ doped XC8-0 cylinders, including a bidirectional CDW phase (see SM for details) emerges around $t'=0$ and z-stripe phase for $t'>0.5$. }
    \label{fig4:zigzag}
\end{figure}

{\it Geometry-stabilized CDW orders:}
The lattice geometry of finite cylinders imposes significant constraints on stripe formation \cite{Xu2024Pair}. For instance, the a-stripe configuration depicted in Fig.~\ref{fig2:SCcorandcharge}(b) is geometrically frustrated on both XC8-0 and YC4-4 cylinders. This geometric incompatibility leads to the suppression of a-stripe and promotes competing states with translational invariant pattern along the periodic direction. To systematically probe this geometry dependence, we further investigate the ground states of the $t$-$t'$-$J$-$J'$ models on XC8-0 and YC4-4 cylinders. 

On $1/8$ doped XC8-0 cylinders, DMRG simulation reveals several phases with distinct density profiles. At $t'=0$, we find a bidirectional CDW phase which breaks translation symmetries along both $\hat{e}_x$ and $\hat{e}_y$ directions. This state, however, become unstable upon introducing $t'$, driving the system into a phase separation regime within $t'\sim0.2 - 0.5$ in Fig.~\ref{fig4:zigzag}(c) (See SM for detailed evolution of density patterns). Further increasing $t'$ induces a new phase with pronounced zigzag-oriented stripes (z-stripe). In contrast to the quasi-long-range a-stripe CDW order in YC4-0 cylinder, the amplitude of density oscillation shown in Fig.~\ref{fig4:zigzag}(a) barely changes in the bulk of the cylinders, indicating a long-range CDW order. On the contrary, the long-distance SC correlations in the z-stripe phase exhibit a exponential decay $\Phi(r)\sim e^{-r/\xi_{sc}}$ with a short correlation length $\xi_{sc}\sim 0.4$ unit cell, suggesting the absence of superconductivity.

On YC4-4 cylinder with $t'=0$ and $\delta=1/12$, we find another type of z-stripe state characterized by coexisting quasi-long-range SC and CDW, fully consistent with prior study of the $t'=0$ Hubbard model \cite{Peng2025} on YC4-4 cylinder. However, the increasing $t'$ frustrates this stripe order, driving the system into phase separation regime for $t'\sim 0.1 - 1.0$ (see SM for details).

{\it Slave boson mean-field theory:} The emergence of geometry-dependent competing phases naturally prompt the question of what constitutes the true ground state in the 2D thermodynamic limit, where the lattice geometry effects become negligible. Restricted by the entanglement entropy grown with system width, DMRG can not directly access wider systems. Therefore, we employ SBMFT calculations to evaluate the fate of all candidate CDW states identified in DMRG simulations as the systems approach to 2D limit. 

% Figure 5: MF phase diagram
\begin{figure}[tb]
    \centering
    \includegraphics[width=1.0\linewidth]{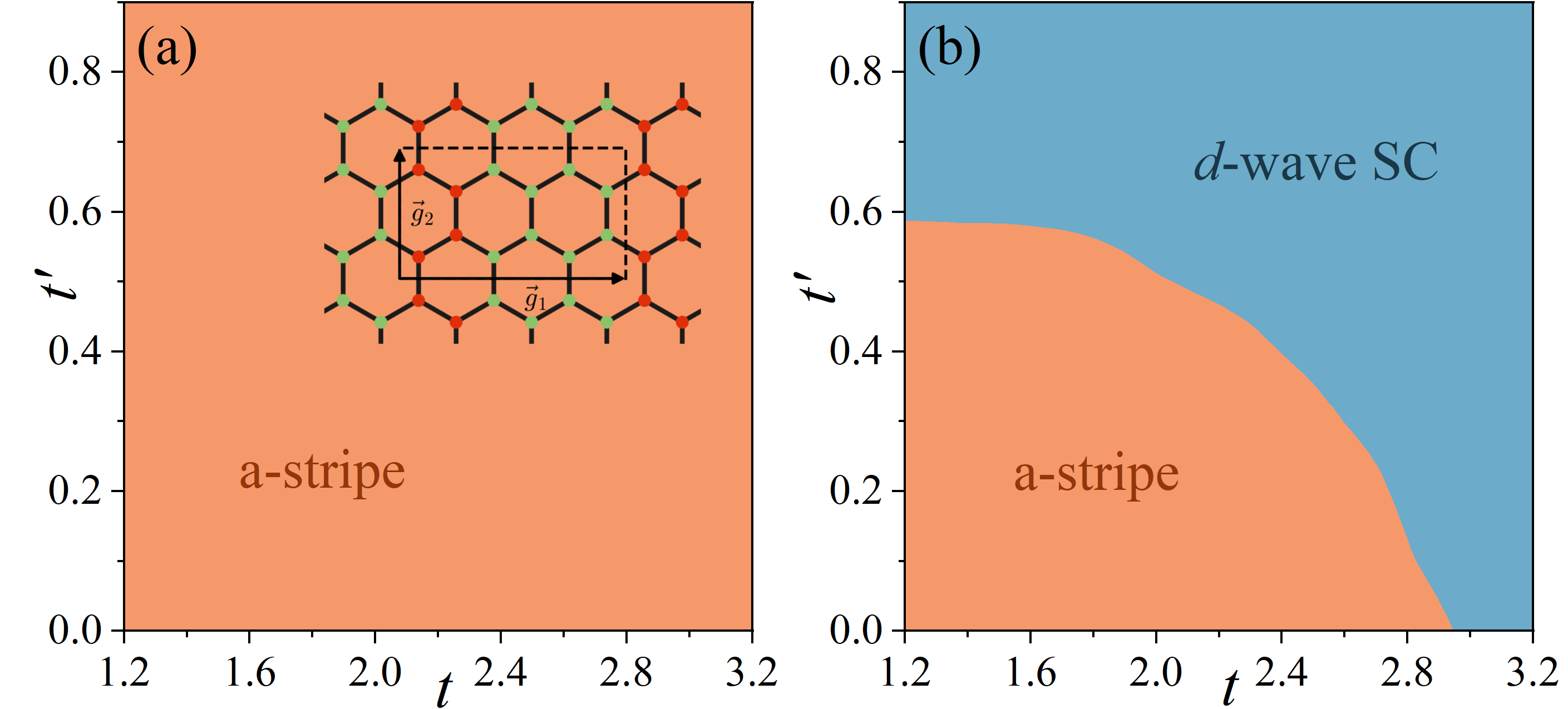}
    \caption{SBMFT phase diagrams for 2D $t$-$t'$-$J$-$J'$ models with $J=1$. (a) $\delta = 1/12$ case: a-stripe phase dominates over other competing states across the entire $t$-$t'$ region we tried. (b) $\delta = 1/8$ case: a nematic $d$-wave SC phase with uniform charge density (blue region) emerge for large $t$ and $t'$. 
    Inset: the enlarge unit cell and primitive vectors of the a-stripe order.}
    \label{fig5:MF-phase}
\end{figure}

In the SBMFT framework, the electron creation operator is decomposed as $c^\dagger_{i\sigma}=b_i f^\dagger_{i\sigma}$, where $b_i$ and $f_{i\sigma}$ denote the bosonic holon and fermionic spinon annihilation operators, respectively. The Hamiltonian is decoupled in both particle-hole and particle-particle channels with local constraint $b_i^\dagger b_i + \sum_{\sigma} f^\dagger_{i\sigma} f_{i\sigma}=1$ introduced by Lagrange multiplier $\lambda_i$.
In order to capture all CDW candidates, the order parameters $\chi_{ij}=\left <\sum_{\sigma}f^\dagger_{i\sigma} f_{j\sigma}\right>$, $\Delta_{ij}=\left < f^\dagger_{i\uparrow} f^\dagger_{j\downarrow}-f^\dagger_{i\downarrow} f^\dagger_{j\uparrow}\right>$ and $\delta_{ij}=\left< b_i^\dagger b_j\right>$ are defined in the enlarged unit cells commensurate with the selected CDW pattern (e.g., the a-stripe unit cell in inset of Fig.~\ref{fig5:MF-phase}). 
The mean-field ground state is determined by comparing converged energy obtained from self-consistent calculations across a series of initial density distribution $n_{f,i}=\chi_{ii}$ mimicking different CDW candidates (see SM for details). Since the establishment of CDW strongly suppresses SC ordering at the MF level, our MF analysis primarily focuses on the dominant CDW pattern among candidates. SC properties are investigated for the state with uniform density profiles.

The MF phase diagrams for $\delta=1/12$ and $1/8$ models with a range of $t$ and $t'$ $(J=1)$ are illustrated in Fig.~\ref{fig5:MF-phase}.
For $\delta=1/12$, the a-stripe state consistently yields the lowest energy across the entire parameter region, dominating over the z-stripe and other competing CDW states in 2D limit. At $\delta=1/8$, the a-stripe remains dominant in small $t'$ region. 
However, a transition from a-stripe to nematic $d$-wave SC phase with uniform charge density occurs when $t'$ cross a critical value ($\sim 0.5$ for $t=2.0$) in Fig.~\ref{fig5:MF-phase}(b). This transition reasonably aligns with  DMRG observations on YC4-0 cylinders, where the charge correlations weaken significantly while the SC pair correlation remains dominant for $t'>0.5$ regime in Fig.~\ref{fig3:exponent}(b).

{\it Conclusion and discussion: } In this work, we have presented a comprehensive study of the extended $t$-$t'$-$J$-$J'$ model on the honeycomb lattice. By employing large-scale DMRG simulations and SBMFT, we elucidate the profound role of $t'$ in enhancing SC and its intricate competition with CDW order.
On YC4-0 cylinders, DMRG reveals a robust SC phase with sub-leading CDW at $\delta=1/12$ and $1/8$ doping. The Luttinger exponent $K_{sc}$ of this phase exhibits a valley-shaped non-monotonic dependence on $t'$, with an optimal SC enhancement occurred at $t'_{op} \sim 0.4$. This large SC phase suggests that the pairing mechanism in doped honeycomb model may be more robust and less reliant on the specific type of undoped phase such as QSL.
Moreover, the corresponding ``optimal'' $J' \sim 0.04$ lies deep within the AFM regime of the undoped phase diagram, indicating that the SC enhancement could be primarily driven by fluctuation within the doped AFM state rather than the quantum critical points or QSL.
Given the tunable realization of Hubbard physics in twisted Moiré systems and their multicomponent extensions \cite{Yuan2018, Pan2020, Angeli2021, Pan2023, Ma2025, Wei2025}, our results may stimulate efforts to search unconventional superconductivity in the corresponding $t'\sim t'_{op}$ regime.

Another important finding of our work is the significant dependence of the emergent low-energy phases on the lattice geometry of finite cylinders. While DMRG simulations on YC4-0 cylinders reveal a robust SC phase across a wide range of $t'$ and doping, distinct states emerge under the same model parameters on XC8-0 and YC4-4 cylinders. This pronounced geometry dependence underscores the challenge of extrapolating finite-width cylinder results to 2D. However, we can still take advantage of this sensitivity by treating it as a new tuning parameter that tipping the balance between the competing orders. By properly choosing or designing boundary condition of finite-size lattices, we can effectively probe different facets of the complex competition between the low-energy orders to gain deeper insights into the underlying physics. Similar strategy is recently applied to square-lattice Hubbard model and successfully reveals a pair-density-wave SC coexisting with charge stripes \cite{Xu2024Pair}. 
The synergy of DMRG and SBMFT, which combining state-of-the-art simulations on quasi-1D systems with complementary theoretical calculations to probe the 2D limit, provides a general and effective way to navigate complex competing orders in 2D strongly correlated systems.

{\it Acknowledgments: } YFJ is supported in part by the National Key R$\&$D Program of China under Grants No. 2022YFA1402703, NSFC under Grant No. 12347107 and 12574160. H.-C.J. is supported by the US Department of Energy, Office of Basic Energy Sciences, Division of Materials Sciences and Engineering, under Contract No. DE-AC02-76SF00515.

\bibliography{second}

\clearpage

\begin{widetext}

\renewcommand{\theequation}{S\arabic{equation}}
\setcounter{equation}{0}
\renewcommand{\thefigure}{S\arabic{figure}}
\setcounter{figure}{0}
\renewcommand{\thetable}{S\arabic{table}}
\setcounter{table}{0}

%==appendix==
\section{Supplemental Material}

\subsection{Lattice geometry and numerical details}
The three cylinder geometries used in DMRG simulations, YC4-0, YC4-4 and XC8-0, are illustrated in Fig.~\ref{Sfig:geometry}. X and Y indicate the orientation of the lattice bond, for example, 
the YC cylinders have one type of NN bonds, i.e. the $a$ bond in Fig.~\ref{Sfig:geometry}, lying along the $\vec{e}_y$ direction. While for XC cylinders, the orientation of lattices is defined by the $b$ bond lies along the $\vec{e}_x$ direction. The YC$m$-$n$ cylinder consists $m$ zigzag chains along $\vec{e}_x$ direction, with periodic connections shifted by $n$ sites along the zigzag chain. While for XC$m$-$n$ cylinders, $m$ denotes the number of sites on the zigzag chain along $\vec{e}_x$ and $n$ is the number of shifted unit cells at periodic boundary. 

The length of cylinder $L_x$ is defined by the number of two-site unit cell along the $\vec{e}_x$ direction. 
In this work, we mainly investigate the property of models on four-leg systems with $L_x=48$, which is long enough to avoid boundary effect in the investigated parameter range. In most of the calculations, the number of DMRG block states is kept up to $m=24000$ to achieve a truncation error $\epsilon \lesssim 8 \times 10^{-6}$ for each models.

\begin{figure}[h]
    \centering
    \includegraphics[width=0.8\linewidth]{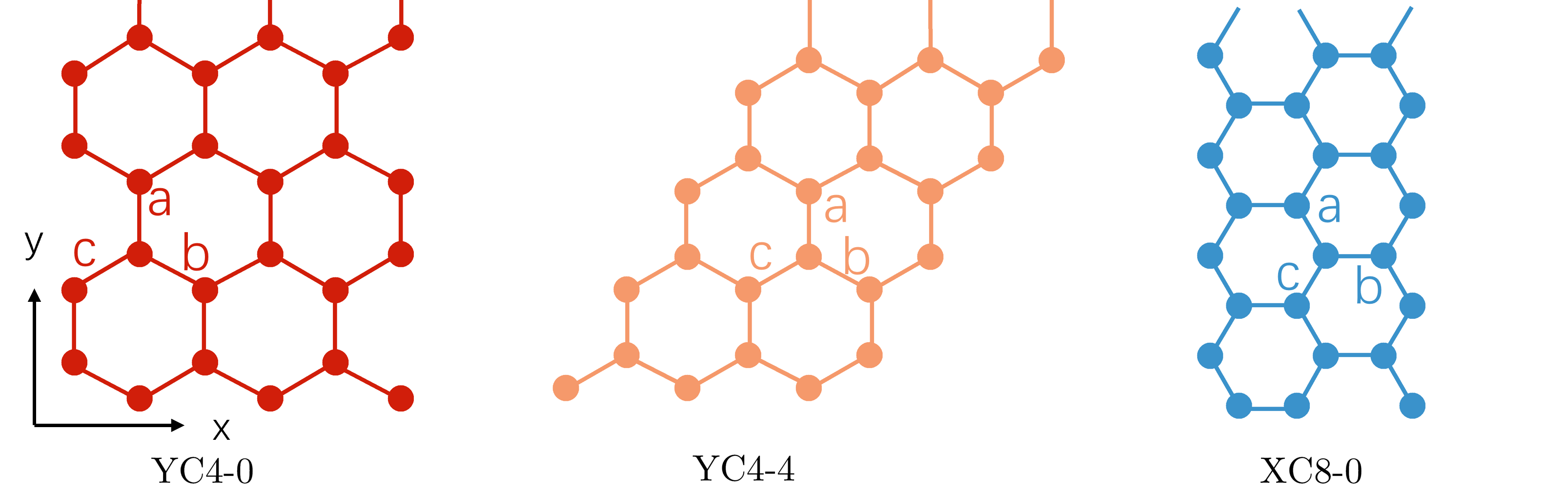}
    \caption{The three different geometries for four-leg cylinders with length $L_x=3$. $a$, $b$ and $c$ denote the types of bonds.}
    \label{Sfig:geometry}
\end{figure}

We perform finite truncation error extrapolation to reduce the truncation error in our DMRG simulations with finite bond dimensions. The second-order polynomial $O(\epsilon)=O_0+a_1\epsilon+a_2\epsilon^2$ is employed to extract the converged physical quantity $O_0$ in the zero truncation error limit $\epsilon \rightarrow 0$, based on the results $O(\epsilon)$ measured with $m=15000$ to $24000$ kept states. The extrapolation for both pair-pair correlation functions $\Phi(r)$ and rung-average charge densities $n(x)$ for $t'=0.4$ models on $N=2 \times 48 \times 4$ YC4-0 cylinder with $\delta=1/12$ and $1/8$ are illustrated in Fig.~\ref{Sfig:scaling}. For the pair-pair correlation function $\Phi(r)$, extrapolations are independently applied to the data at each spatial separation $r$ to extract entire correlation functions in the zero truncation-error limit. 
In Fig.~\ref{Sfig:scaling}(a) and (b), we illustrate the curves of representative results for $\Phi(r=14\sqrt{3})$ in $\delta=1/12$ and $1/8$ models. 
The fitting curves of rung average charge density $n(x)$ at site $x=18\sqrt{3}$ are illustrated in Fig.~\ref{Sfig:scaling}(c) and (d). 

\begin{figure}[h]
    \centering
    \includegraphics[width=0.95\linewidth]{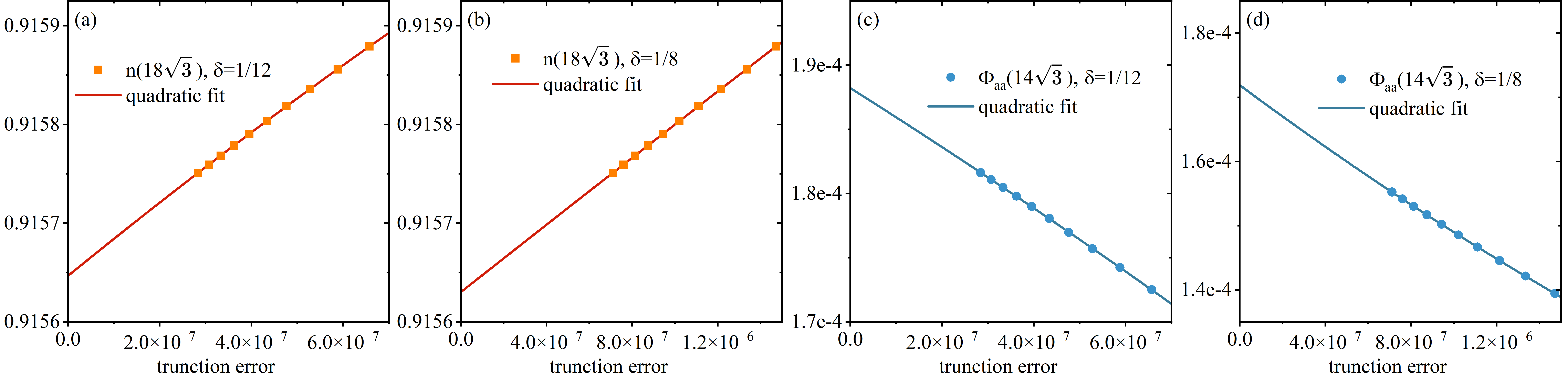}
    \caption{The finite-truncation-error extrapolations for rung-average charge densities and the pair-pair correlation functions: (a) the extrapolation of rung average charge density $n(x=18\sqrt{3})$ for $\delta=1/12$ case and (b) for  $\delta=1/8$ case. (c) the extrapolation of long-distance pair-pair correlation $\Phi(r=14\sqrt{3})$ for $\delta=1/12$ case and (d) for $\delta=1/8$ case.}
    \label{Sfig:scaling}
\end{figure}

\subsection{Properties of $\delta=1/8$ doped model on YC4-0 cylinders}
Here we show that the SC phase observed in the $\delta = 1/12$ model on YC4-0 cylinders in the main text remains robust at the higher doping $\delta = 1/8$. Following the same procedure, we show in the Fig.~\ref{Sfig:charge_sc_8} that the detailed properties of $\delta = 1/8$ doped model are nearly identical to those of the $\delta = 1/12$ case. The SC correlations decays in power law with decay exponent $K_{sc}$ around $0.79 \sim 0.97$, exhibiting a valley-shape dependence on $t'$. The relative magnitudes of the SC correlations satisfy $\Phi_{aa}(r) \sim 4 \Phi_{bb}(r) \sim 4 \Phi_{cc}(r)$. 
The wavelength of the stripe is $\lambda = \sqrt{3}/(4\delta) = 2\sqrt{3}$ (i.e., $1/4\delta$ unit cells) along the $\hat{e}_x$ direction.
As discussed in the main text, at $\delta = 1/8$ doping the CDW decay exponent monotonically increases with $t'$, indicating a strong suppression of the CDW order at larger $t'$ for $\delta = 1/8$ models. This is in good agreement with the slave-boson mean field results.

\begin{figure}[h]
    \centering
    \includegraphics[width=0.95\linewidth]{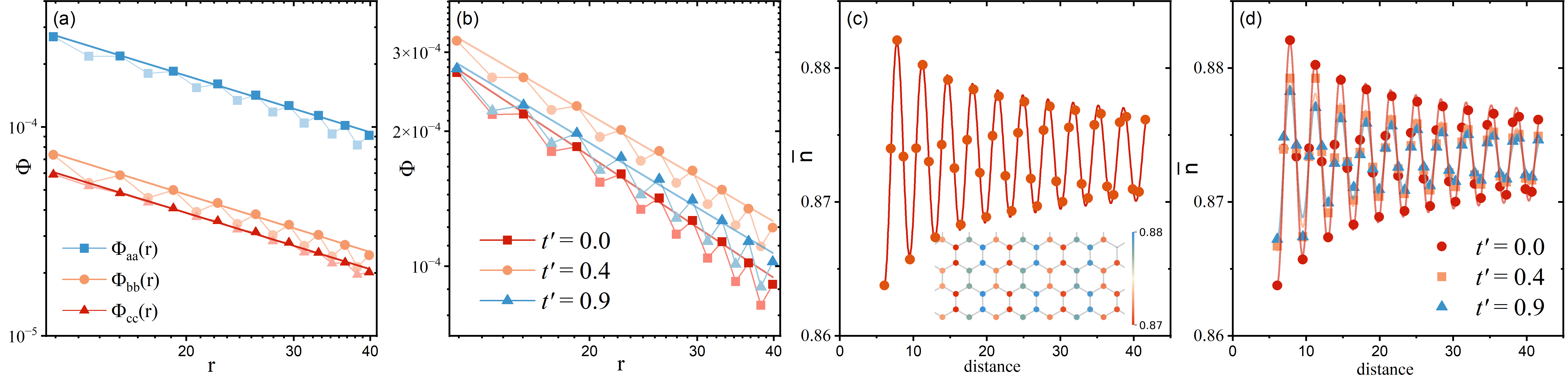}
    \caption{The SC and charge properties of $\delta=1/8$ doped model on YC4-0 cylinders: (a) Three types of SC correlations $\Phi_{aa}$, $\Phi_{bb}$ and $\Phi_{cc}$ obtained at $t'=0$, fitted to power-law functions $f(r)\sim r^{-K_{sc}}$ (solid line). The transparent points far from the envelope are discarded in the fitting process. (b) The SC correlation $\Phi_{aa}$ measured at $t'=0.0$, $0.4$ and $0.9$. (c) The rung-averaged charge density $n(x)$ of the $t'=0$ model, fitted to the Friedel oscillation (solid line). Inset: Charge density profile in the bulk. (d) The charge density oscillations obtained from same models in (c).}
    \label{Sfig:charge_sc_8}
\end{figure}

\begin{figure}[h]
    \centering
    \includegraphics[width=0.85\linewidth]{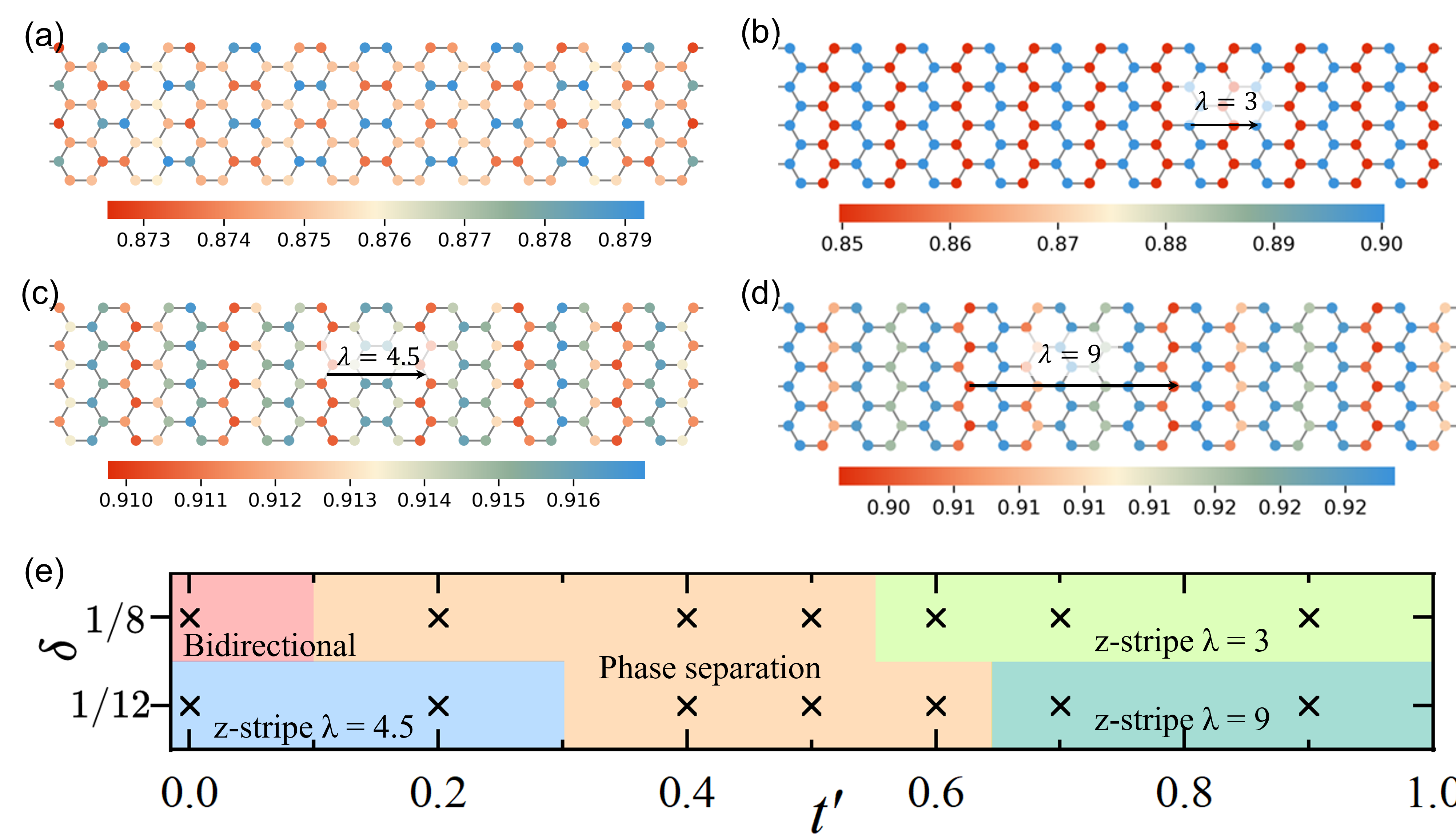}
    \caption{The charge patterns and the quantum phase diagram for models on the XC8-0 cylinder: (a) the bidirectional CDW pattern in the $\delta = 1/8$ and $t'= 0.0$ model. (b) The long-range z-stripe order with wavelength $\lambda = 3$ in the $\delta = 1/8$ and $t'=0.6$ model. (c) The z-stripe order with wavelength $\lambda = 4.5$ in the $\delta = 1/12$ and $t'= 0.0$ model. (d) The z-stripe order with longer wavelength $\lambda = 9$ observed in the $\delta = 1/12$ and $t'= 0.9$ model. (e) The quantum phase diagram for the models on XC8-0 cylinder.}
    \label{Sfig:XC8-0}
\end{figure}

\subsection{$t'$ phase diagram on XC8-0 cylinders}

Fig.~\ref{Sfig:XC8-0} illustrates the phase diagram of the model on XC8-0 cylinders and representative charge patterns in each phases.
At $\delta = 1/8$, the system exhibits a bidirectional CDW pattern around $t' = 0$ (see Fig.~\ref{Sfig:XC8-0}(a)) and a strong z-stripe phase with wavelength $\lambda=3$ in $t' > 0.5$ regime (Fig.~\ref{Sfig:XC8-0}(b)). As discussed in the main text, the charge-density oscillation in this z-stripe phase does not decay even in the middle of cylinders, indicating a long-range stripe orders. Between these two CDW phases, the system is phase separated.

For $\delta = 1/12$ case, the system develops a long-range z-stripe phase with a wavelength $\lambda = 4.5$ for $t' = 0\sim 0.2$, as shown in Fig.~\ref{Sfig:XC8-0}(c). Similar to the $\delta = 1/8$ case, this z-stripe phase becomes unstable as $t'$ increases, and the system enters phase-separated regime consists of two or more different density patterns. In the large $t'>0.6$ regime, we observe another long-range z-stripe phase with longer wavelength $\lambda=9$ developed. The SC order is absent in whole phase diagram on XC cylinders.

\subsection{Properties of models on YC4-4 cylinders}
For the YC4-4 cylinder, the charge density profile of the $t'=0$ and $\delta = 1/12$ model develops another type of z-stripe pattern with a wavelength $\lambda = \sqrt{3} /{4\delta}$ along $\vec{e}_x$ direction as depicted in Fig.~\ref{Sfig:YC4-4}(a). The SC correlations of this model exhibit power-law decay with exponent $K_{\mathrm{sc}} \sim 1.1$ and the amplitude of three types of correlations satisfy $\Phi_{aa}(r)\sim \Phi_{cc}(r) \gg \Phi_{bb}(r)$ shown in Fig.~\ref{Sfig:YC4-4}(c). This result is consistent with previous DMRG studies of the honeycomb-lattice Hubbard model \cite{Peng2025}.
However, the z-stripe pattern is not stable in $t'=0$ model with slightly higher doping $\delta = 1/8$. As illustrated in Fig.~\ref{Sfig:YC4-4}(b), a bidirectional CDW with density pattern similar to that in Fig.~\ref{Sfig:XC8-0}(a) develops on YC cylinders. This bidirectional CDW order competes strongly with the z-stripe order for $t'>0$ models on YC4-4 cylinders, inducing phase separations in most of the $t'$ we studied.

\begin{figure}[h]
    \centering
    \includegraphics[width=0.85\linewidth]{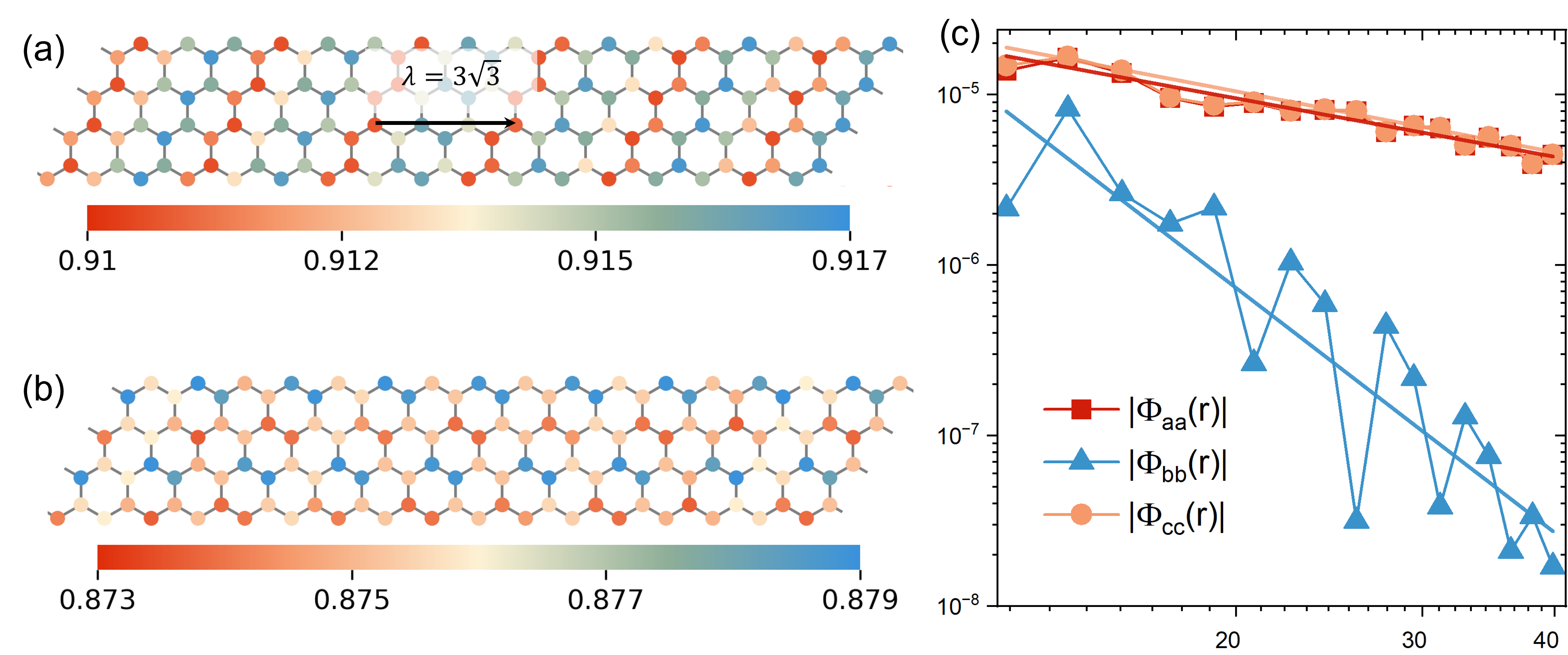}
        \caption{The properties of $t'=0$ models on the YC4-4 cylinder. (a) The charge density profile showing another type of quasi-long-range z-stripes for the $t'= 0.0$ model with $\delta = 1/12$. (b) The bidirectional CDW pattern developed in the $t'= 0.0$ model with slightly higher doping $\delta = 1/8$. (c) The three types of SC correlations measured in the $t'= 0.0$ model with $\delta = 1/12$. The leading correlations $\Phi_{aa}$ and $\Phi_{cc}$ exhibit quasi-long-range behavior with a power-law decay exponent $K_{sc}\sim1.1$. }
    \label{Sfig:YC4-4}
\end{figure}

\subsection{Slave boson mean field theory}
We employ the slave-boson mean-field approach to calculate the possible charge patterns of the doped honeycomb-lattice $t$-$t'$-$J$-$J'$ model in 2D limit. All the stable charge orders observed in the DMRG simulation have been consider as candidate states in mean field calculation. Using the slave-boson approach, we decompose the electron into the fermionic spinon and bosonic holon $c_{i\sigma}^\dagger = b_i f_{i\sigma}^\dagger$ with constraint $b_{i}^{\dagger}b_i+\sum_{\sigma}{f_{i\sigma}^{\dagger}f_{i\sigma}}=1 $ and rewrite the Hamiltonian as
\begin{eqnarray}
	H=&&-t_{ij}\sum_{ij\sigma}{b_if_{i\sigma}^{\dagger}f_{j\sigma}b_{j}^{\dagger}} + h.c.\nonumber\\
	&&-\frac{J_{ij}}{2}b_i b_j b_j^{\dagger} b_i^{\dagger}\sum_{ij\sigma}{\left( f_{i\sigma}^{\dagger}f_{j-\sigma}^{\dagger}f_{j-\sigma}f_{i\sigma}+f_{i\sigma}^{\dagger}f_{j-\sigma}^{\dagger}f_{i-\sigma}f_{j\sigma} \right)} , 
\end{eqnarray}
where the spin interaction $J_{ij}b_i b_j b_j^{\dagger} b_i^{\dagger}$ is further approximate as $J_{ij}(1+\delta_i)(1+\delta_j) \sim J_{ij}$ at small doping. The mean field decoupling of this Hamiltonian is applied in both particle-hole and particle-particle channels 

\begin{eqnarray}
	H_{MF}=&&-\sum_{ij\sigma}{\left( t_{ij}\delta _{ij}^{\dagger}+\frac{1}{4}J_{ij}\chi _{ij}^{\dagger} \right) f_{i\sigma}^{\dagger}f_{j\sigma} }-\sum_{ij}{t_{ij}\chi _{ij}^{\dagger}b_{i}^{\dagger}b_j+h.c. } \nonumber \\
	&&-\sum_{i\sigma}{\left( \sum_{j\in \left( i,j \right)}\frac{1}{4}{J_{ij}n _{j}}+\mu +\lambda _i \right) f_{i\sigma}^{\dagger}f_{i\sigma}}\nonumber \\
	&&-\sum_{ij}{\frac{1}{2}J_{ij}\left[ \Delta _{ij}^{\dagger}(f_{i\downarrow}f_{j\uparrow}-f_{i\uparrow}f_{j\downarrow})+h.c. \right]} + const. ,
    \label{EqS2:HMF}
\end{eqnarray}
where order parameters $\chi_{ij} = \sum_{\sigma}{\langle f_{i\sigma}^{\dagger}f_{j\sigma} \rangle}$ denotes the spinon hopping amplitude for $i\ne j$ and local fermion number $n_i=\chi_{ii}$ for $i=j$. $\delta_{ij} =\langle b_{i}^{\dagger}b_j \rangle $ represent the holon hopping amplitude, and $\Delta_{ij} = \langle f_{i\downarrow}f_{j\uparrow}-f_{i\uparrow}f_{j\downarrow} \rangle$ the spin-singlet pairing amplitude. $\lambda_i$ is the Lagrange multiplier for the local constraint $b_{i}^{\dagger}b_i+\sum_{\sigma}{f_{i\sigma}^{\dagger}f_{i\sigma}}=1$. The total number of fermions is adjusted by the spinon chemical potential $\mu_f$, and holons are condensed in the lowest-energy boson state for zero temperature calculation. The constant term in Eq.~\ref{EqS2:HMF} takes the form $\sum_{ij\delta}{t_{ij}\delta _{ij}^{\dagger}\chi _{ij}}+h.c.+\sum_{ij\sigma}{\frac{1}{4}J_{ij}}\left( |\chi _{ij}|^2+n_in_j \right) +\sum_{ij}{\frac{1}{2}J_{ij}|\Delta _{ij}|^2}$.

\begin{figure}[h]
    \centering
    \includegraphics[width=0.9\linewidth]{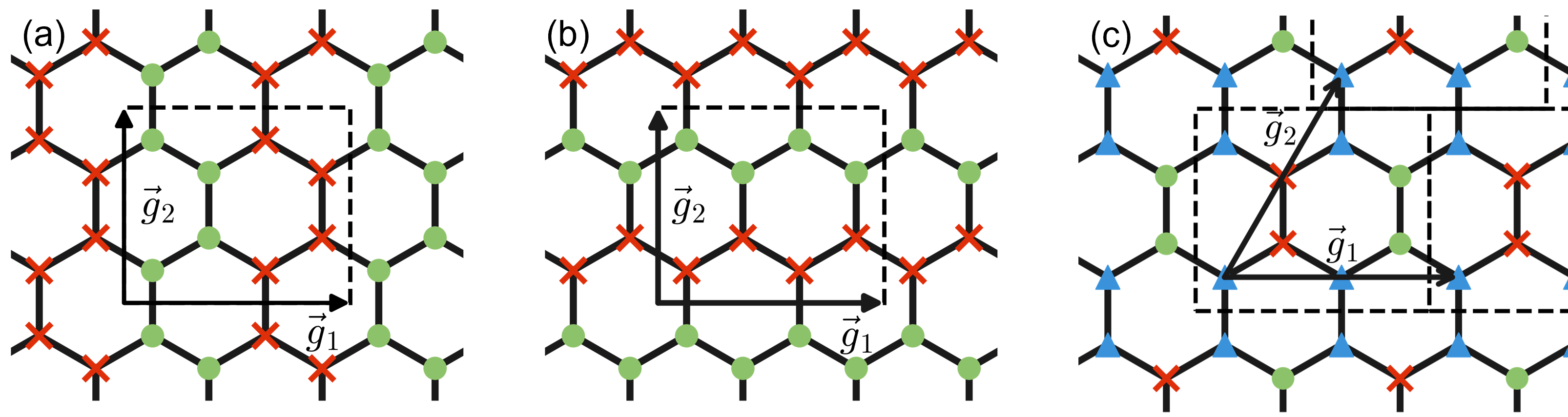}
    \caption{Schematic diagram of different patterns: (a) a-stripes, (b) z-stripes, and (c) bidirectional CDW. Different colors denote different charge densities.}
    \label{Sfig:MF}
\end{figure}

We mimic the charge density patterns observed in DMRG calculation by properly adjusting the order parameters $n_i$ within enlarged CDW unit cells in the initial step of MF self-consistent iterations. Examples of enlarged unit-cell configurations for a-stripes, z-stripes and bidirectional CDW for $\delta=1/8$ are shown in Fig.\ref{Sfig:MF}. To explicitly enforce these candidate charge patterns, averaging of charge densities across symmetry-equivalent sites is systematically applied during each mean-field iteration. The resulting phase diagram is determined by identifying the lowest-energy state from a series of optimized solutions obtained with distinct initial charge patterns.

\end{widetext}

\end{document}